# Ionic Sieving Through One-Atom-Thick 2D Material Enables Analog Nonvolatile Memory for Neuromorphic Computing

Revannath Dnyandeo Nikam[1,2,*], Jongwon Lee[1,2], Wooseok Choi[1,2], Writam Banerjee[1,2], Myonghoon Kwak[1,2], Manoj Yadav[1,2], Hyunsang Hwang[1,2,*]

[1]Center for Single Atom-based Semiconductor Device, [2]Department of Material Science and Engineering, Pohang University of Science and Technology (POSTECH), Pohang 790-784, Republic of Korea

* *Authors to whom correspondence should be addressed*

*Email: revannikam@postech.ac.kr; hwanghs@postech.ac.kr*

## Abstract

The first report on ion transport through atomic sieves of atomically-thin 2D material is provided to solve critical limitations of electrochemical random-access memory (ECRAM) devices. Conventional ECRAMs have random and localized ion migration paths; as a result, the analog switching efficiency is inadequate to perform in-memory logic operations. Herein ion transport path scaled down to the one-atom-thick (~0.33 nm) hexagonal boron nitride (hBN), and the ionic transport area is confined to a small pore (~0.3 nm$^2$) at the single-hexagonal ring. One atom thick hBN has ion-permeable pores at the center of each hexagonal ring due to weakened electron cloud and highly polarized BN bond. Our experimental evidence indicates that the activation energy barrier for $H^+$ ion transport through single-layer hBN is approximately 0.51 eV. Benefiting from the controlled ionic sieving through single-layer hBN, the ECRAM exhibit superior non-volatile analog switching with good memory retention and high endurance. The proposed approach enables atomically-thin 2D material as an ion transport layer to regulate the switching of various ECRAM devices for artificial synaptic electronics.

**Keywords:** Hexagonal boron nitride; 2D material; ionic transport; artificial synapses; neuromorphic computing

## 1. Introduction

Electrochemical random access memory (ECRAM) can emulate synaptic functions for neuromorphic computing.[1] In ECRAM, a gate dielectric is replaced with an ion-conducting gate electrolyte that includes dissolve mobile ions (e.g; $H^+$, $Li^+$, $O^{2-}$).[1b, 2] Analog switching in the channel of ECRAM occurs when mobile ions from gate electrolyte are pumped into or out of the ion-sensitive channel. The ionic mobility and stability of ion diffusion pathway in the gate electrolyte determines the analog switching characteristics of ECRAMs.[1a, 3] To achieve linear switching in ECRAM devices, modern material engineering advancements are speeding up the development of electrolyte materials with lower activation energy barriers and channel materials with lower redox energy barriers.[1a, 4] Proton based ECRAM (H-ECRAM) is potentially advantageous because proton ($H^+$) is a smaller and more rapidly diffusing ion than $Li^+$ and $O^{2-}$. However, the memory state-retention times of H-ECRAM far remained relatively short with cycling instability.[5] Due to H-ECRAM' poor memory state-retention, inference accuracy degrades over time. The poor retention in H-ECRAM is caused by self-discharge of intercalated ions from the channel layer due to exitance of non-zero open circuit potential (OCP), which is mainly occurred in polymer-based H-ECRAM.[6] When the gate circuit opens during the read operation, the electrically conductive electrolyte is unable to halt the backflow of electrons, resulting in non-zero OCP and subsequent self-discharge of $H^+$ ions. Fuller et al. connected a selector in series with the H-ECRAM gate terminal to address the OCP issue, forming a one-selector-one-H-ECRAM (1S1E) structure that isolates the device and prevents leakage current.[7] In our opinion, the memory state-retention and cycling stability of H-ECRAM can be improved by developing proton-conducting solid electrolytes with electron-blocking properties to lower the self-discharge issue. However, the unavailability of a CMOS-compatible proton-conducting solid electrolyte is the main obstacle. All H-ECRAM presently relied on electrolytes that either cannot be integrated and scale down, such as polymer, [8] ionic liquid,[9] ionic gel,[10] organic material.[5] Herein, atomically thin single-layer hexagonal boron nitride (hBN) is integrated into H-ECRAM as a proton-conducting solid-state electrolyte. Atomically thin 2D material has not yet been exploited in prior research for the purpose of improving memory state-retention and cycling stability of ECRAM devices. Recent research has proven that a few two-dimensional (2D) materials exhibit ion transport properties both experimentally and theoretically.[11] Hexagonal boron nitride (hBN) single-layers have been evaluated as a possible material for developing novel ionic transport layers.[12] The honeycomb structure of 2D h-BN is composed of alternating boron and nitrogen atoms.[13] It exhibits superior chemical and thermal stability, as well as mechanical strength.[14] Proton conduction through monolayers of hBN and graphene has been demonstrated theoretically and experimentally.[11c, 12a, 15] Due to the different electron cloud densities of B, N, and C atoms in hBN and graphene, proton permeable pores form in the centers of hexagonal rings.[12a] However, 2D semiconductors like $MoS_2$ are poor proton conductors due to dense electron clouds and a lack of ion-permeable pores.[12a]. The activation energy to $H^+$ permeation through mechanically exfoliated crystals is 1.34 eV for graphene and 0.7 eV for single-layer hBN.[12a] The thin electron cloud in the one-atom-thick hBN layer within B-N hexagon is the main transport medium for fast $H^+$ ions. Additionally, the electrically insulating nature of hBN and ~0.33-nm ion transport path yields energy-efficient operation. The strategy of ion transport through atomically thin 2D material sheds new light on ECRAM device design, paving the way for the rapid construction of artificial synapses for a high-sensitivity neuromorphic computing system. By implementing the ECRAM with single-layer hBN, we established the physical limit for vertical thickness scaling of the ion transport layer (0.33 nm).

## 2. Results and Discussion

### 2.1. 2D hBN Enables High-Performance ECRAMs and Working Principle

To investigate how $H^+$ transport through hBN affects synaptic characteristics, ECRAM devices were fabricated and tested under DC double sweep and AC pulse measurements. The single-layer hBN was sandwiched between a tungsten trioxide ($WO_3$) channel and silicon -proton (Si-H) reservoir; the resulting devices are denoted as hBN-ECRAM. The switching mechanism of hBN-ECRAM is schematically illustrated in **Figure 1A-C**. Initially, in the absence of an electric field at gate terminal (i.e., gate voltage $V_G$ = 0 V) the hBN-ECRAM is in its native high-resistance state (HRS), in which no net movement of $H^+$ ion occurs (**Figure 1A**). After applying a small positive $V_G$, $H^+$ ions migrate from their reservoir to $WO_3$ through the single-layer hBN (**Figure 1B**). The B-N bonding in hBN is highly polarized due to higher electronegative N atom; studies suggest that the higher electronegative nitrogen atom tends to pull more electrons, which results in the formation of a weak electron cloud with ion-permeable pore (~3 $Å^2$) at the center of hexagonal ring.[11c, 12a, 16] Small cations ($H^+$ and $Li^+$) can easily pass through the negatively-charged electron cloud.[15, 17] With single-layer hBN, the transport of $H^+$ ion occurs at a constant rate that may result in uniform $WO_3$ channel switching with high endurance and good retention. The transport of $H^+$ ions from the Si proton reservoir to the $WO_3$ channel results in $H_xWO_3$ conducting phase, and the device reaches the low resistance state (LRS) with a lower $V_{set}$ of 1.71 V (**Figure 1C**). To provide additional evidence for the development of $H_xWO_3$, we examined the oxidation/reduction states shift in gated and ungated $WO_3$ channels using XPS measurements; see Supporting Information **Figure S1**. For XPS analysis, we used back gate ECRAM devices to directly trace the phase change in $WO_3$. The direct exposer of the XPS beam on the $WO_3$ channel in the back gate ECRAM can eliminate hBN and Si-H interference during XPS analysis. Without gate bias (Vg = 0 V), the appearance of ($W^{+6}$) doublet peaks at high binding energy indicates the presence of the $WO_3$ phase. Positive gate bias (Vg =1 V) results in the formation of $H_xWO_3$ phase, as evidenced by the presence of ($W^{+5}$) doublet peaks at lower binding energy **(Figure S1).** To test our hypothesis, we quantified how switching and transfer characteristics depend on $V_G$. We modulated the channel current $I_{SD}$ by cycling the $V_G$ from 0 to 3 V, 3 to 0, and 0 to -3 V at a rate of 10 mV $s^{-1}$. In hBN-ECRAM, channel current increases as $V_G$ increases positively (**Figure 1D**); this response indicates transport of $H^+$ ion from Si proton reservoir to $WO_3$ channel. $H^+$ intercalation in $WO_3$ results in the formation of $H_xWO_3$ conducting phase. Without hBN, $I_{SD}$ increased to a certain value, but immediately returned to the original state in the same gate polarity when $V_G$ was withdrawn, see Supporting Information **Figure S2**. The high volatility of $I_{SD}$ is attributed to the significant gate leakage in ECRAM devices that do not contain hBN. In the absence of hBN, the ECRAM device exhibits switching due to electron-hole transport, which results in small hysteresis windows and highly volatile conductance states. The high volatility in the channel causes several shortcomings in synapse devices, including small memory windows, poor ON/OFF ratio, and non-identical conductance change $\Delta G$. To ensure high performance and efficient learning at the hardware level, the ideal synapse system should have 7 bits non-volatile multi states with a broad dynamic range of 10.[18] Controlled ionic sieving through atomically-thin 2D material may enable these characteristics, and for this purpose, hBN is a good option due to its electrical insulating property, ~0.33 nm thickness, and weak electron cloud within the hexagonal lattice. To test this hypothesis, single-layer hBN was introduced between $WO_3$ channel and Si proton reservoir, then transfer characteristics were measured **(Figure 1D)**. Switching showed counterclockwise hysteresis. As $V_G$ increased positively, the channel current increased to certain value and remained stable even after $V_G$ was withdrawn. Channel-conductance change induced by positive $V_G$ can be erased by applying negative $V_G$ with small $V_{reset}$ of -0.72 V. The polarity-dependent switching behavior in the hBN-ECRAM device and the large and stable hysteresis windows of $I_{SD}$ under $V_G$ indicate that switching in the $WO_3$ channel is non-volatile; this characteristic can be useful to emulate synaptic functionalities. The power consumption in our proposed hBN-ECRAM devices is an important factor to be considered when fabricating neural networks that operate at low power. The power consumption in ionic switchable devices is composed mainly of switching power and leakage power. Gate leakage current $I_G$ is several orders smaller than $I_{SD}$, which can contribute to low-energy operation in hBN-ECRAM devices (**Figure 1D**).

In practical applications, synaptic devices are expected to respond to fast pre-synaptic training pulses. To confirm the fast training and memorizing compatibility of hBN-ECRAM device, several pre-synaptic voltage pulses (1 V, 100 ms) were supplied, then post-synaptic current was measured at low $V_{SD}$ = 0.1 V. Channel conductance increased from $G_1$ to $G_2$ during application of $V_G$, then remain stable to $G_1$ when the pulses were removed (**Figure 1E**); a small transient peak indicates small gate leakage (~ 8 pA) due to presence of the hBN layer. In comparison to hBN, the SiH layer exhibits a larger gate leakage current of 13 nA at 0.5 V. (**Figure S3**). The hBN-ECRAM device showed a good response to pre-synaptic voltage pulses and showed distinct conductance change. The highly stable conductance state in the channel results from zero net movements of $H^+$ after removal of $V_G$. Both discrete conductance and stable conductance change ($\Delta G$) ensure the mimicking of synaptic activities, such as linear switching, good retention, and endurance. In contrast, without hBN, the channel conductance increased suddenly following gate pulses, followed by a rapid decline to the initial state (see Supporting Information **Figure S4**). Furthermore, the absence of a hBN layer results in a strong transient peak, indicating substantial gate leakage. The thickness and surface morphology of ion-conducting layer ultimately govern the ion-conduction path.[3] In this connection, the morphology of the hBN surface was examined using atomic force microscopy (AFM). Scanning AFM data suggested a smooth, homogeneous surface and estimated the thickness of hBN to be 0.41nm (**Figure 1F**).

The elemental stoichiometry of $WO_3$ and hBN was investigated experimentally using X-ray photoelectron spectroscopy (XPS). The XPS spectrum of W 4f (**Figure 1G**) shows two prominent peaks at 35.9 eV (W $4f_{7/2}$) and 38.0 eV (W $4f_{5/2}$) that confirm the presence of a nearly stoichiometric $WO_3$ phase. In **Figure 1H-I**, the B1s peaks occur at 190.4 eV, and the N1s peaks occur at 398.1 eV, which are consistent with a hexagonal boron nitride phase.[19] The Raman spectrum of hBN shows the presence of the $E_{2g}$ phonon mode at 1365 $cm^{-1}$ and full-width at half-maximum (FWHM) of 14.5 $cm^{-1}$, confirming its single-layer nature **(Figure 1J)**.[13a]

### 2.2. Wafer-Scale 2D hBN Transfer Process Enable Scalable ECRAM Assembly

The use of wafer-scale synaptic devices is emerging to fulfill the need to produce practical neuromorphic applications. Therefore, we demonstrated a device-scale-to-wafer fabrication process for hBN-ECRAM (**Figure 2**). The fabrication process starts with the deposition of $WO_3$ on a 4-inch $SiO_2$/Si wafer. To fabricate the various dimensions of the channel photolithographic patterns were used, followed by $WO_3$ deposition and lift-off processes (**Figure 2A**). As-fabricated channels were electrically interconnected using Pt source/drain (S/D) metal contact deposition **(Figure 2B**). Large-scale single-layer hBN was obtained using chemical vapor deposition (CVD), then transferred onto the prefabricated $WO_3$ channel pattern **(Figure 2C**). Transfer of the large-scale 2D material is the bottleneck in 2D material device scaling and integration.[20] The interface in the ECRAM device governs the ionic transport mechanism,[21] so to overcome the drawbacks of previously reported small-scale wet-transfer processes,[22] a clean transfer process must be developed. Thus, achieving the large-scale clean transfer process of the 2D material ionic transport layer in ECRAM device is an important step. Therefore, we developed an inexpensive dry-transfer process for hBN (Experimental section). Rolling of hBN-thermorelease tape on the $WO_3$ channel pattern and subsequent heat treatment to release the thermal tape yielded a clean heterostack of hBN-$WO_3$ (**Figure 2D**). Then the hBN-ECRAM was fabricated by depositing Si and top Pt gate metal contact electrodes (**Figure 2E**). Optical microscopy (OM) images (**Figure 2F**) of as-fabricated hBN-ECRAM devices using our developed clean transferred process show that they were fabricated with a fixed $WO_3$ channel width is set at 10 µm and channel lengths ranging from 5 µm to 100 µm. **Figure 2G** depicts a photograph of our scalable and cost-effective dry transfer process, which can transfer a 5 cm single-layer hBN film to a four-inch $SiO_2$/Si wafer. To confirm the effectiveness of our developed dry-transfer process for ECRAM, high resolution cross-sectional transmission electron microscope (HR-TEM) was performed (**Figure 2H**). The HR-TEM image clearly shows the sandwich of single-layer hBN between $WO_3$ and proton doped Si reservoir with a sharp interface. Incorporating a neuromorphic framework into robotics is important to create a diverse range of stretchable and flexible synaptic devices. For flexible electronics, 2D materials such as hBN possess high tensile strength and mechanical flexibility. A prototype ECRAM device was fabricated on PET using flexible hBN (**Figure 2I**).

### 2.3. Proton ($H^+$) Transport Through 2D hBN

When two or more layers of 2D material are stacked, the electron clouds at the centers of their hexagonal lattices can flip independently. Therefore, single-layer hBN can be expected to transport $H^+$ with a lower activation energy barrier, and this ability may improve the analog switching characteristics in the ECRAM devices. To test our hypothesis that $H^+$ transport depends on hBN thickness, ionic tunneling current measurements were performed using vertical Pt/hBN/Pt two-terminal devices (**Figure 3A**). Taken together, our observations establish that single-layer hBN contributes high ionic tunneling current due to the transport of large number of $H^+$ through an atomically-thin electron cloud. The ionic tunneling current in decreased asymptotically as the stacking of the hBN layer increased from double-layer to multilayer. The transport of $H^+$ from the top Pt electrode to the bottom electrode through suspended hBN results in a shift in ionic tunneling current (top inset **Figure 3A**). Prior to measuring the ionic tunneling current, atomic hydrogen ($H^{\cdot\cdot}$) was first implanted into the top Pt electrode using a "spillover" procedure,[23], and then split into a proton and an electron under electric bias. The proton is then transported via hBN to the bottom electrode, and the electron is transported via an external circuit, resulting in a shift in the current level. Our experimental observation clearly shows that the weak electron cloud in single-layer hBN is the main path for $H^+$ transport (bottom inset **Figure 3A**). The variation in current flow characteristics is replotted on a log scale to visualize large discrepancies of values on a single axis (**Figure 3B**). To measure the proton transport, hBN were stacked on Pt hole structure (**Figure 3C**). A schematic of a hBN-stacked hole structure is used to monitor proton transport. The hBN is suspended between the bottom and top electrodes with a ~nm gap. Since the ~nm gap allows the measurement of the contribution in the current shift by ionic species motion rather than electronic charge. Cross-sectional TEM image of hole structure. $SiO_2$ sidewalls height of 100 nm and a hole diameter of 250 nm (**Figure 3D**). To predict the energy barrier to transport $H^+$ through single-layer hBN, temperature dependent conductance measurement was performed (**Figure 3E**), which are found to exhibit the Arrhenius-type behavior, $G \propto \exp(-E/kT)$. Previous experimental analysis predicted that the activation energy barrier of mechanically exfoliated single-layer hBN for $H^+$ is $0.45 \pm 0.04$ eV.[11b] Our experimentally determined activation energy barrier value was ~0.51 eV, which agrees with w the reported value. The slight difference may occur because of moisture ($H_2O$), as well as $H^+$ in Pt-H. With the addition of bilayer and multilayer hBN, $H^+$ transport became more complicated due to the closed pack electron cloud, resulting in an increase in the activation energy barrier to 0.75 eV and 0.98 eV, respectively. To evaluate this atomic-level interpretation, HR-TEM analysis was performed on the single-layer hBN in the hole structure device (**Figure 3F**). The hBN single-layer was predicted to be ~ 0.33 nm thick without any mechanical damage region, to ensure only $H^+$ transport through the pores in the electron cloud. The HR-TEM analysis also considered an Pt/hBN/Pt device that used stacked double-layer hBN (**Figure 3G**). The double-layer hBN was

~0.68 nm thick and the layers were distinct. The negligible switching in Pt/hBN/Pt that use multi-layer hBN was also supported by HR-TEM analysis (**Figure 3H**).

## 2.4. Synaptic and Neuromorphic Functions in hBN-ECRAM

To confirm the synaptic characteristics, ECRAM device that used single-layer hBN were tested using identical AC pulse schemes (**Figure 4A**) that alternate 'write' and 'read' operations. The 'write-erase' operation were performed using a high-amplitude pulse (± 1 V, $t_{width}$ = 10 ms, $t_{interval}$ =10 ms) and subsequent 'read' operation were performed using small pulses (0.1 V, 10 ms). Analog switching in hBN-ECRAM device was programmed by alternatively applying total of 148 identical positive and negative 'write-erase' gate pulses. In hBN-ECRAM, analog conductance switching by discrete conductance states with linear and symmetrical conductance transition was achieved (**Figure 4B**). Long-term potentiation (LTP) and long-term depression (LTD), which are both important synaptic plasticity phenomena, are closely resembled by the increasing and decreasing channel conductance. Linear LTP and LTD operations in a synapse device are highly desirable for achieving efficient learning capability in an artificial neural network. The atomically thin (0.333 nm) ionic transport path through single-layer hBN, which is more controllable than ion transport through a traditional bulk electrolyte, is responsible for the linear LTP and LTD switching in hBN-ECRAM. Furthermore, the linear analog switching in hBN-ECRAM device was understood by monitoring a region of channel resistance change using current sensing atomic force microscopy (CS-AFM) in response to $H^+$ transport from gate (**Figure S5**)). To perform the CS-CAFM analysis, the two sets of back gated ECRAM devices with and without hBN were fabricated. The back gated ECRAM device allows us to scan the $WO_3$ surface directly with the CS-AFM tip (**Figure S5 (A-B**). Prior to CS-AFM measurements, both ECRAM devices was set to LRS by applying 0 to 3 V sweeping bias to gate terminal. Then the change in resistance of $WO_3$ due to the formation of $H_xWO_3$ phase was mapped using CS-AFM at the channel region (**Figure S5 (C-F**). The current mapping result on hBN lack ECRAM shows the small centralized region of $WO_3$ channel switches to LRS while the rest of the region remains at HRS (**Figure S5 (C-D**). Contrarily, sieving of $H^+$ through hBN delivering the de-centralized switching in large region of $WO_3$ channel (**Figure S5 (E-F)**). The decentralized switching in $WO_3$ is attributed to the uniform distribution of $H^+$ ions through hBN, which results in consistent ΔG with linear-analog switching throughout the operation (**Figure 4 (C)**). The learning ability of synaptic devices depends on analog switching capabilities such as dynamic range, multilevel memory states, and bidirectional linear switching. As a result of control and consistent $H^+$ transport through single-layer hBN, a dynamic range of 1.3 with consistent memory change $\Delta G$ ~ 0.06 µS was obtained. The multilevel memory states in a single synapse allow storage of large numbers of bits, and can thereby achieve efficient arbitrary programming operations. The present hBN-ECRAM synapse shows 64 (= $2^6$; 6 bits) conductance levels. Such high-bit operation in a single ECRAM device results from uniform $H^+$ distribution inside the $WO_3$ channel and low electrical gate leakage disturbance through single-layer hBN.

Linearity in potentiation and depression is critical for evaluating synapse weight update action. To quantify the linearity, a nonlinearity parameter ($\alpha$) was determined using the previously described methods.[24] The extracted nonlinearity under identical pulses in the hBN-ECRAM parameters were $\alpha_p$ = 0.90 and $\alpha_d$ = 0.90. The switching linearity of the hBN-ECRAM device was extremely desirable; nonlinearity values for potential and depression are one in an ideal case (i.e., $\alpha_p = \alpha_d = 1$). The hBN-ECRAM device real-time non-volatile characteristics (**Figure 4C**). Nonvolatility is an important factor for both on-chip and off-chip training to retain the memory state in synaptic devices.[25] To test the nonvolatility, single write pulses of ± 1 V with amplitude 100 ms were supplied to the gate terminal, then the change in channel conductance was measured for 100 s with a 0.1-V pulse. During potentiation and depression, each channel conductance increased to a certain value and remained stable for 100 s with $\Delta G$ ~0.1 µS. These results indicate excellent non-volatility. Additionally, analog synaptic activities were established in a transparent and flexible hBN-ECRAM system; see Supporting Information **Figure S6**. In synaptic devices, cycle-to-cycle variation directly affects the recognition accuracy of neuromorphic computing.[18] Therefore, this variation should be small. Cyclic endurance measurements were performed in hBN-ECRAM **(Figure 4D)**. To test endurance up to 58400 pulses, a high-speed pulse scheme (± 1 V, $t_{width}$ = 10 ms, $t_{interval}$ =10 ms) was applied to the gate terminal. After $10^4$ cycles, $G_{\max}$ had dropped by < 5%; therefore, the devices have excellent endurance for training the Modified National Institute of Standard and Technology (MNIST) data set. The $\Delta G$ uniformity during cycling operation influences the accuracy of the neural networks and ideally, should have a narrow distribution. The conductance variance ($\Delta$G) density distribution in hBN-ECRAM was obtained for 58400 cycles (**Figure 4E-F**). Switching accuracy was 98 % for potentiation cycling and 97 % for depression cycling.

In neural networks, the ability of synaptic devices to retain information is a critical factor for the inference process.  A neural network model trained using supervised learning is an inference process that can be used to make predictions against previously unseen data.[18] For the test of retention characteristics, three conductance states (8.9 µS, 9.9 µS, 10.9 µS) were chosen, and their distribution over $10^3$ s was measured. To achieve the desired level of $WO_3$ channel conductance, a series of gate voltage pulses were applied to the gate electrolyte terminal. The voltage on the gate electrolyte terminal was then removed, and the $WO_3$ channel current was measured by applying a 0.1-V DC bias to the S/D terminal. The selected conductance state degraded negligibly for up to $10^3$ s (**Figure 4G**). The excellent retention characteristics of the hBN-ECRAM system are due to the long-term stability of $H^+$ in the $WO_3$ channel over the course of gate operation. It is worth noting that, despite the fact that hBN-ECRAM has lower memory retention characteristics than two terminal memristor, the overall neuromorphic performance remains unchanged, as memory retention is more important in memory storage applications than in neuromorphic applications. In a neuromorphic chip, on-chip training is frequently completed in < 1 h. As a result, synapse devices are not subjected to 10-year retention mandates.

The high speed of synaptic devices is a crucial characteristic, particularly for online training of neural networks.[25-26] The switching speed of the hBN-ECRAM system was measured using different gate pulses with varying pulse duration (**Figure 4H**). The hBN-ECRAM device shows distinct ΔG with 10-ms pulse. A noticeable improvement in channel conductance was obtained by increasing the width of the gate pulse (50, 100, 500 ms). This pulse width-dependent switching characteristic determines the switching speed of the hBN-ECRAM. The current hBN-ECRAM device switches 10 ms faster than previously-reported synaptic devices that use electrolyte $H^+$. The high speed of switching in the present ECRAM was achieved because of the short ~0.33 nm ionic path in the single-layer hBN. The synaptic device variability can influence the classification accuracy of a neural network.[27] Therefore, device-to-device conductance distribution for maximum / minimum conductance states (8.36 and 10.53 µS) was measured in 30 hBN-ECRAM devices (**Figure 4I**). All 30 devices showed narrow conductance distributions of 0.04 µS during potentiation and 0.005 µS during the depression and so had excellent device-to-device uniformity. Precision in terms of repeatability and reproducibility is critical to validate the findings. To ensure reproducibility, seven batches of experiments were carried out under identical conditions to fabricate the hBN-ECRAM devices. Interestingly, all devices from seven experiments shows the same switching behavior. The reproducibility in seven experiments is statistically represented using conductance states distribution at $10^{th}$, $20^{th}$, $40^{th}$, $80^{th}$, $100^{th,}$ and $140^{th}$ pulse numbers (**Figure S7**).

## 2.5. Neuromorphic Computing Simulation on hBN-ECRAM Device Parameters

To demonstrate the use of hBN-ECRAM in pattern recognition, a multi-level cell (MLC) neural network was simulated using the potentiation and depression experiment data (**Figure 4D**). For the simulations, the MLC was composed of 528 input neurons as a 1st layer, 250 +125 hidden neurons as a 2nd layer, and 10 output neurons as a 3rd layer (**Figure 5A)**. Simulations are considered an ideal synapse composed of a ECRAM device with < 50 levels of multistate, linearity ($\alpha_p = \alpha_d = 1$), and large dynamic range in conductance change. A completely connected neural network is referred to as an MLC, and in this type of network, each neuron node in one layer is linked to any neuron node in the following layer. Synapses are associations between neurons, and the strength of the connection can be dynamically adjusted. MNIST input image is given by 528 input neurons (i.e. the number of input neurons that correspond to a 24x22 pixel image), and each input neuron is connected to one pixel of the image. Each of the ten output neurons is linked to one of the ten-digit groups (0-9). Each neuron acts as a single perceptron. The 1st hidden layer receives 528 inputs, and the 2nd -hidden layer collects 528 inputs. The collected input signals are multiplied by the synaptic weight represented by conductance change; then the signals are accumulated by a transfer function. The accumulated result is modified by a sigmoid activation function and transferred to the input of the next layer. The MLC was trained on 20000 patterns at each epoch, which were chosen at random from 60,000 images in the MNIST database. The results of the recognition accuracy test were then compared to the results from the separate dataset, which included a separate collection of 10,000 images. **Figure 5B** depicts a mathematical model of an MLC neural network with signal processing and transmission workflow for a single training cycle. The synaptic weights ($w$) of each synapse correspond to the hBN-ECRAM channel conductance. Activations ($x$) of each neuron are propagated to the next neuron via the non-linear function (f) during the forward propagation process. In the simulations, the fabricated hBN-ECRAM achieved a recognition accuracy of 93.27% after 20 training epochs, comparable to the 95.71% obtained using an ideal synapse (**Figure 5C**). In addition, it was also observed that the image recognition accuracy of the hBN-ECRAM was equivalent to that of an ideal synapse (**Figure 5D**). The ideal synapse, in this case, has 10-bit of multi-states and perfect linearity ($\alpha_{p,d}$ =1). The high recognition accuracy in hBN-ECRAM is attributed to the > 50 multi-states and reasonably good linearity ($\alpha_p$ = 0.90; $\alpha_p$ = 0.90).

## 3. Conclusion

In practice, the most significant challenge in ECRAM devices is to balance ion transport in the electrolyte and electronic state change in the channel to achieve near-ideal synaptic characteristics. Many efforts have been made to optimize the electrolyte to improve the synapse characteristics in ECRAM, by the devices had low reliability due to uncontrolled and localized ion migration by exiting the bulk ion transport layer.[1a, 3] In our opinion, reliability in the ECRAM device is a major problem, and can be addressed by reducing the ion-transport path at an atomic scale to realize delocalized single-ion switching. This study has shown that the ion-transport layer can be scaled down to single-atom thickness (~0.33 nm) using single-layer hBN to achieve an ECRAM electronic synapse for ideal neuromorphic systems. To the best of our knowledge, scaling down the ionic transport path to single-atom thickness has not been reported previously. Our innovative exploitation of proton sieving through one-atom-thick hBN greatly improved the synaptic properties of ECRAM compared to reported synaptic devices (**Table 1**). All these observations indicate that 2D materials can be more useful than rigid bulk electrolyte materials as an ion transport layer. These results provide a method to develop of novel 2D material for controlled ionic transport for nanoelectronics and ionic devices.

## 4. Experimental Section

*Device Fabrication*: The 20 µm (L) × 10 µm (W) device H-ECRAM device that used single-layer hBN was fabricated on 100 nm $SiO_2$/Si substrate. Photolithography was used to make the channel, source-drain, and gate patterns. A positive photoresist (PR; AZ GXR 601, Micro Chemicals, Germany) was spin-coated on the Si/$SiO_2$ substrate at 4000 rpm for 30 seconds to create the lithographic patterns. The sample was then baked for 90 seconds on a hot plate at 110 °C. Following that, photolithography was performed using Cr photomask under 365-nm ultraviolet (UV) light for 7 seconds on a MIDAS mask aligner (MDA-400M). The PR was then developed by immersing it for 35 seconds in a developer (AZ-MIF-300, Micro Chemicals, Germany). After that, the sample was rinsed with DI water. A 30 nm thick $WO_3$ was deposited lithographic channel pattern by reactive oxygen sputtering with a 2-in W target. The optimal deposition conditions for $WO_3$ were determined to be T = 25 °C, 100 W plasma power, and a gas flow ratio of 30 sccm Ar/2 sccm $O_2$ at a pressure of 5 mTorr. Afterward, the lift-up process was performed in acetone to form the $WO_3$ channel. The source-drain pattern was then photolithographically pattern on the $WO_3$ channel. Subsequently, 60 nm Pt was deposited on source-drain pattern by DC sputtering using 2-in Pt target under 30 sccm Ar gas flow and 5 mTorr pressure. Afterward, the lift-up process was performed in acetone to form the source-drain contact. To serve as an ion-transport layer, high-quality single-layers of hBN grown on Cu foil by CVD were transferred on $WO_3$ channel as follows. To begin, thermal release tape was used to laminate hBN on Cu foil. The Cu foil was then etched away using a Cu etchant, and the remaining hBN was thermally transferred to a $WO_3$ channel. Then 5 nm Si reservoir and 15 nm Pt electrodes were deposited by sputtering on lithographically prepared gate patterned. To implement $H^+$ in hBN-ECRAM, we used a "spillover" method with Pt electrode as a catalyst, in which $H_2$ molecules from forming gas dissociate to atomic hydrogen, which is converted to $H^+$ by annealing at 100 °C. The $H^+$ is localized into the Si reservoir and can re-distributed under electric bias through hBN and thereby change the resistance of the $WO_3$ channel. The detailed information on the fabrication of Si-H reservoir and its XPS and AFM analysis is supplied in **Figure S8 and Figure S9**, Supporting Information.

*Device and Material Characterization*: Two electrical measurement configurations were used to evaluate the performance of ECRAM devices. First, the AC pulse measurements were performed using the Agilent B1500A semiconductor analyzer with the same instrument's pulse generation unit (PUG). Second, standard double DC sweep measurements using the Agilent B1500A semiconductor analyzer with the three source measurement units (SMUs). X-ray photoelectron spectroscopy (XPS) measurements were used to determine the atomic composition (Thermo VG, U.K. instrument). Atomic force microscopy (Park XE-100 AFM) was used to examine the topologies of single-layer hBN. HR-TEM was performed using a JEOL ARM 200F instrument fitted with an ASCOR probe-corrector in TEM mode.

**Supporting Information**

Supporting Information is available from the Wiley Online Library or from the author.

**Acknowledgments**

This work was supported by the National Research Foundation of Korea (NRF) under Grant 2018R1A3B1052693 and the U.S. Army International Technology Center Pacific under Contract FA5209-20-C-0018.

**Conflict of Interest**

The authors declare no conflict of interest.

**TABLE OF CONTENT**

This study presents the first report of protonic sieving through single-layer hBN to achieve highly reliable analog switching in an electrochemical random-access memory (ECRAM) electronic synapse for an ideal neuromorphic system. Reports shows a practical way to achieve the scaling limit of the ion transport path to atomic thickness (~0.33 nm) by using single-layer hBN. The insight on ionic control through atomic thickness provides guidelines to use 2D materials to replace existing bulk ion transport material.

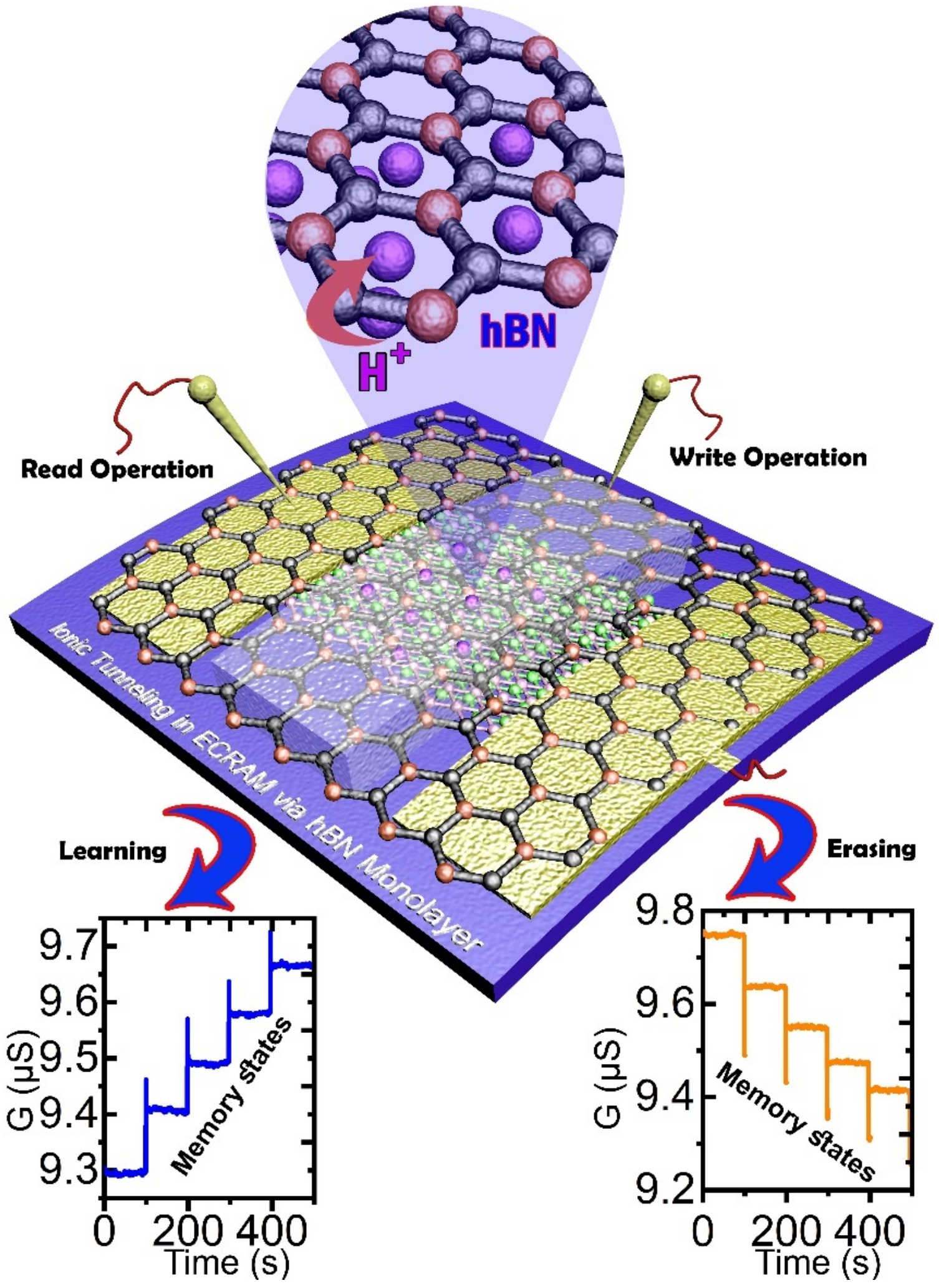

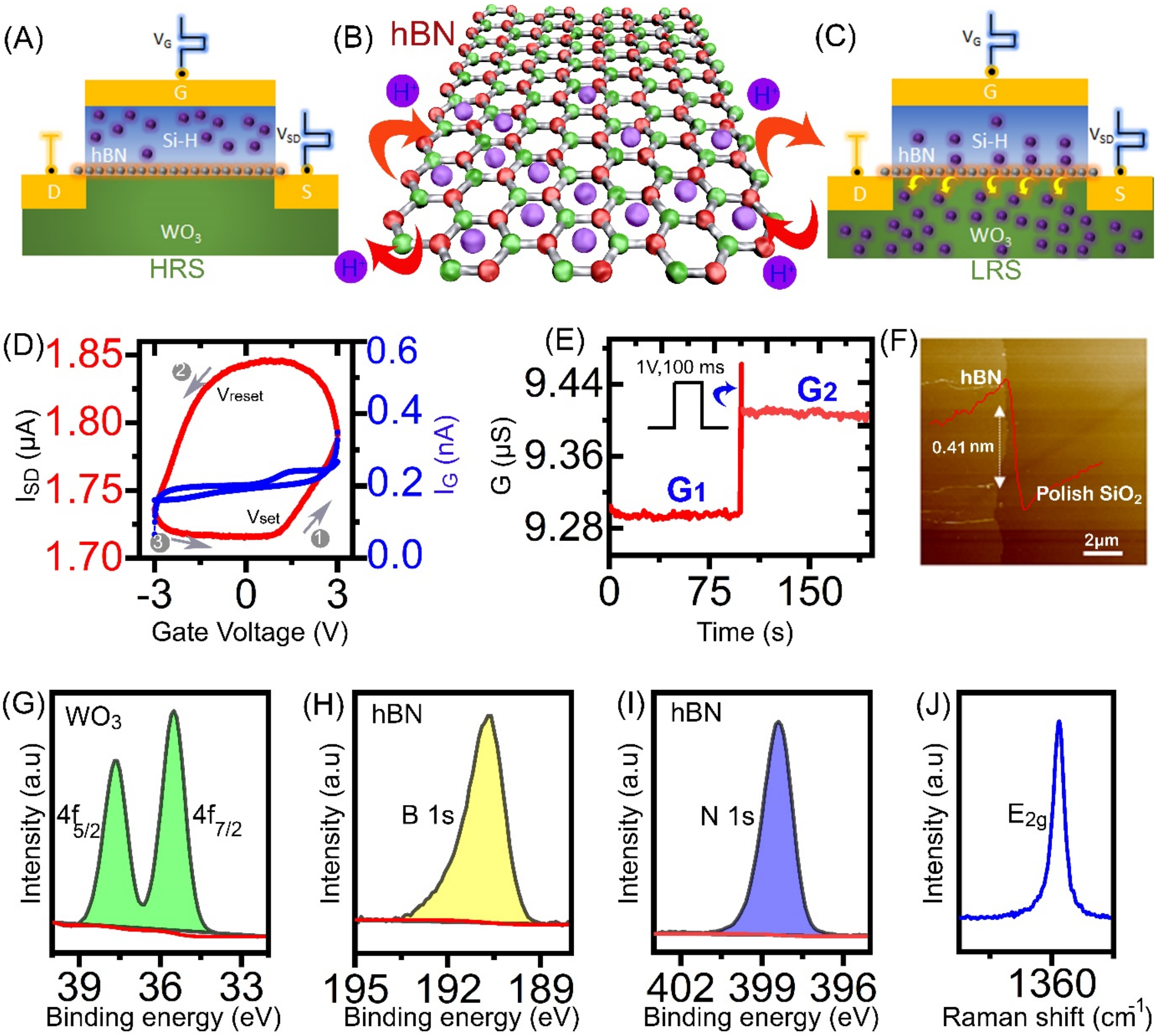

**Figure 1. Effect of $H^+$ ion sieving through single-layer hBN on analog switching in ECRAM. (A)** Schematic illustration of HRS state in the hBN-ECRAM device. At $V_G = 0$, zero net movements of $H^+$ ion into $WO_3$ channel from Si-H through hBN. **(B)** Transport of $H^+$ ion through the weak electron cloud of a hexagonal B-N ring of hBN. **(C)** Schematic illustration of LRS state in the hBN-ECRAM device. Under positive gating, $H^+$ moves to the $WO_3$ channel through single-layer hBN, and the process results in the formation of conducting $H_xWO_3$ phase. **(D)** Transfer characteristics of the hBN-ECRAM device. Numbers: switching sequence; arrows, switching direction. Large hysteresis and bidirectional switching delivering non-volatile memory characteristics to the hBN-ECRAM device. The values of $V_{set}$ and $V_{reset}$ are 1.71 V and -0.72 V, respectively. **(E)** Non-volatile conductance switching in the hBN-ECRAM device. Non-volatile LTP functionality was mimicked by applying high-amplitude presynaptic voltage pulses (1 V, 100 ms) to the Si-H gate terminal through the hBN. **(F)** AFM scan of hBN **(G-I)** Chemical structural analysis of $WO_3$ and hBN using high-resolution core-level XPS. **(J)** Raman spectra of hBN recorded using $\lambda$ = 532 nm excitation laser. The $E_{2g}$ peak (1365 $cm^{-1}$) and 14.5 $cm^{-1}$ FWHM are characteristic of monolayer hBN.

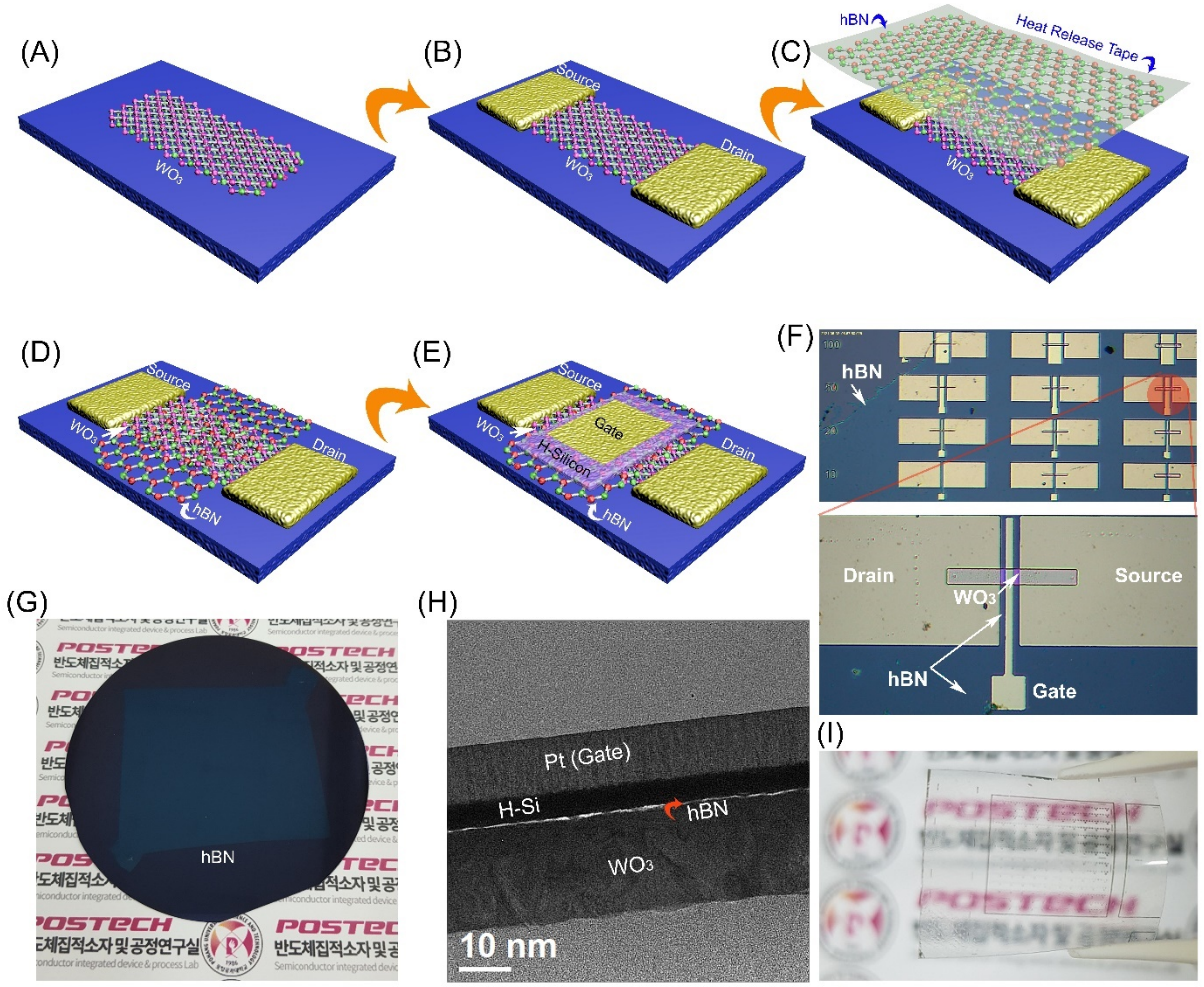

**Figure 2. Fabrication of single-layer hBN consisting ECRAM device. (A)** Patterning of ion-sensitive $WO_3$ switching channel by lithography and magnetron sputtering. Various dimensions (10 ~100 µm) of the channel were deposited to confirm short-channel effects and switching speed. **(B)** Pt metal S/D contact electrodes were deposited on the pattern $WO_3$ channel. **(C)** Single-layer hBN was transferred onto the $WO_3$ channel by a dry-transfer process, which preserves the quality of hBN and $WO_3$ with low levels of contamination. **(D)** Thermal treatment releases the polymeric thermal tape; transfer is effective because of the strong interfacial adhesion between hBN and $WO_3$. **(E)** Si-H layer as an $H^+$ ion reservoir is deposited on selected areas by lithography and magnetron sputtering. The Pt gate metal is deposited, followed by a lift-up process. **(F)** Microscopic photograph of as-prepared hBN-ECRAM device array. Zoom in view indicates the single device.  The zoom-in view is for a single device. A 10-µm channel width was used for all hBN-ECRAM unit, whereas channel length was varied from 5~100 µm × 10 µm. **(G)** Photograph of 5 cm single-layer hBN film transferred to a four-inch $SiO_2$/Si wafer **(H)** Cross-sectional TEM of as-fabricated single-layer hBN-ECRAM. Red arrow: sandwiched layer between $WO_3$ channel and Si-H. The TEM analysis confirms ~0.33 nm thickness of hBN, ~5 nm Si-H and ~30 nm $WO_3$ channel. **(I)** A transparent and flexible hBN-ECRAM device fabricated on PET substrate.

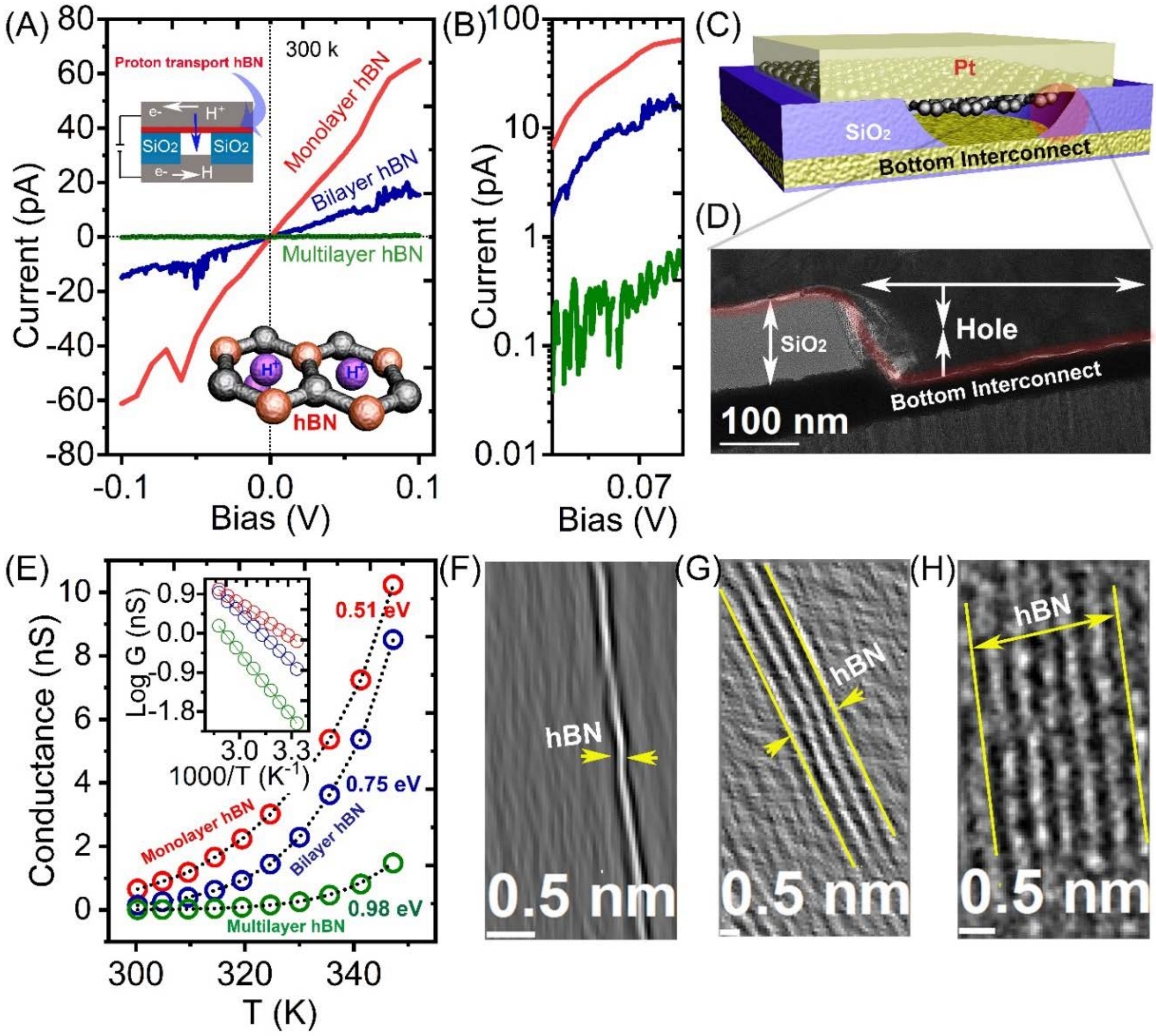

**Figure 3. Ionic sieving and activation energy barrier for $H^+$ ion transport through single-layer, bi-layer, and multi-layer hBN. (A)** The variation in current flow characteristics caused by the extent of $H^+$ ion transport through single-layer, bi-layer, and multi-layer hBN configured two-terminal devices. Top inset: Proton transport schematic in a vertical two-terminal device via hBN layer. Bottom inset: proton transport route scheme in hBN single-layer (red balls representing boron atoms; dark gray, nitrogen; light purple, $H^+$ ion). **(B)** The variation in current flow characteristics is replotted on a log scale to visualize large discrepancies of values on a single axis. **(C)** A schematic of a hBN-stacked hole structure is used to monitor proton transport. The hBN is suspended between the bottom and top electrodes with a ~nm gap. Since the ~nm gap allows the measurement of the contribution in the current shift by ionic species motion rather than electronic charge. **(D)** Cross-sectional TEM image of hole structure. $SiO_2$ sidewalls height of 100 nm and a hole diameter of 250 nm. **(E)** Temperature dependence of conductance shift behavior in monolayer, bilayer, and multilayer hBN devices. The upper inset shows the Arrhenius plot for hBN devices. The solid lines are the best Arrhenius fits ($G \propto \exp(-E/kT)$), and their slopes are used to obtain the activation energy. The activation energies obtained for single-layer, bi-layer, and multi-layer hBN are 0.51, 0.75, and 0.98 eV, respectively. **(F-H)** HR-TEM on monolayer, bilayer, and multilayer hBN. Yellow arrows and lines indicate the direction of the hBN layer.

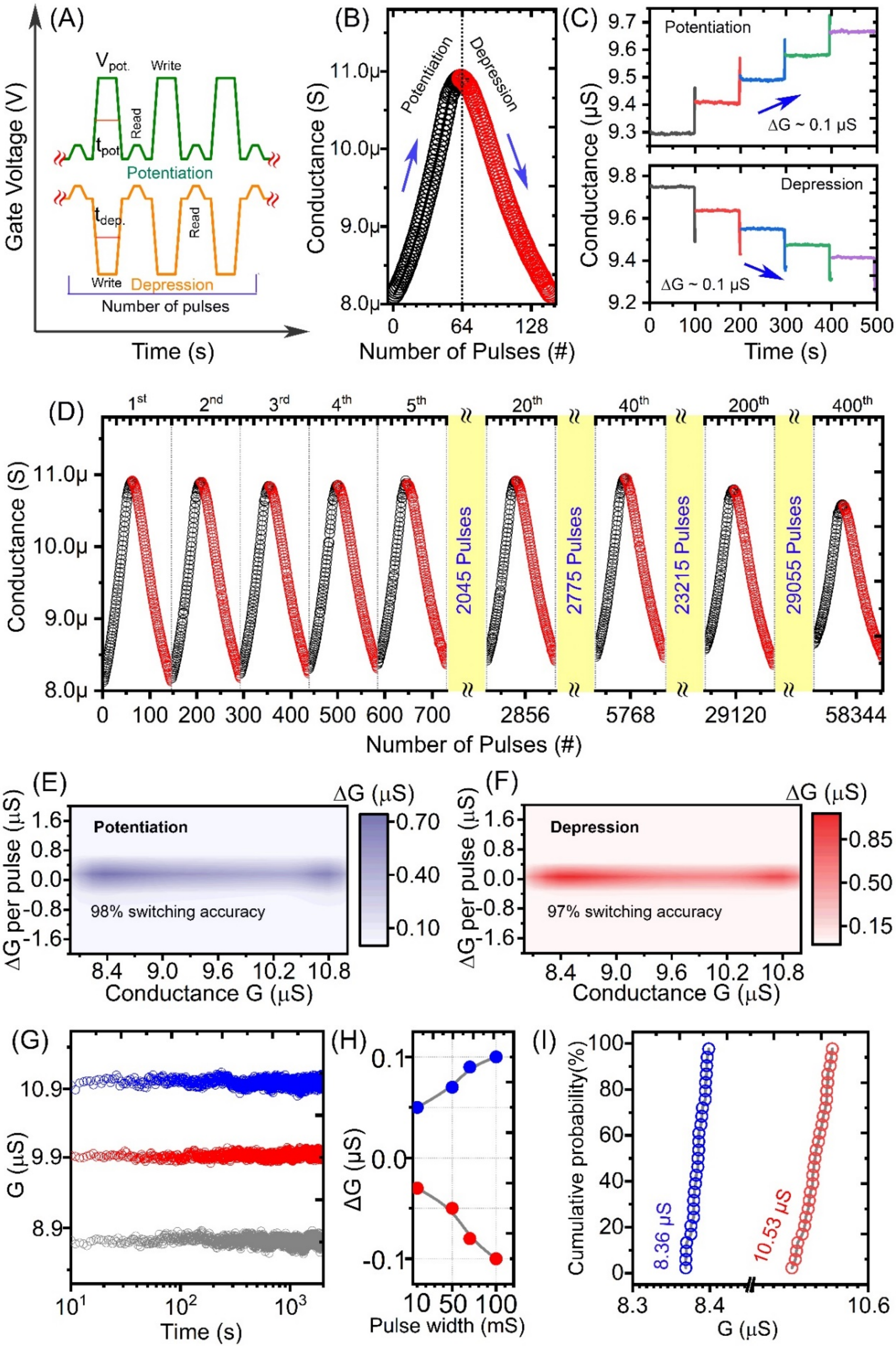

**Figure 4. Analog switching characteristics of single-layer hBN consisting ECRAM synapse. (A)** High-speed AC pulse scheme used to train the hBN-ECRAM synapse. A total of 146 up-down training pulses with amplitude ± 1 V and duration 10 ms were applied to achieve linear and symmetric weight updates in the hBN-ECRAM synapse. For a read operation, AC

pulses with amplitude 0.1 V and width of 10 ms were applied after each training pulse. **(B)** Analog switching behavior in hBN-ECRAM according to the number of training pulses. Linear switching characteristics for both positive (potentiation) and negative (depression) weight updates. **(C)** Non-volatile switching characteristics in the hBN-ECRAM synapse. The pulse-delayed scheme (± 1V,100 ms, space 100 s) is used to realize nonvolatility in adjacent conductance states. **(D)** Endurance characteristics of hBN-ECRAM synapse up to 58400 training pulses with amplitude ±1 V and a duration of 10 ms. **(E-F)** Switching distribution plot of 40 potentiation and depression cycles, respectively. Switching accuracy is predicted by number of training pulses that show uniform conductance change (ΔG). **(G)** Retention characteristics of randomly selected conductance state from hBN-ECRAM synapse. **(H)** Pulse-width-dependent switching speed. The device shows measurable ΔG with the smallest pulse of 10 ms. **(I)** Switching variability test in 30 devices.

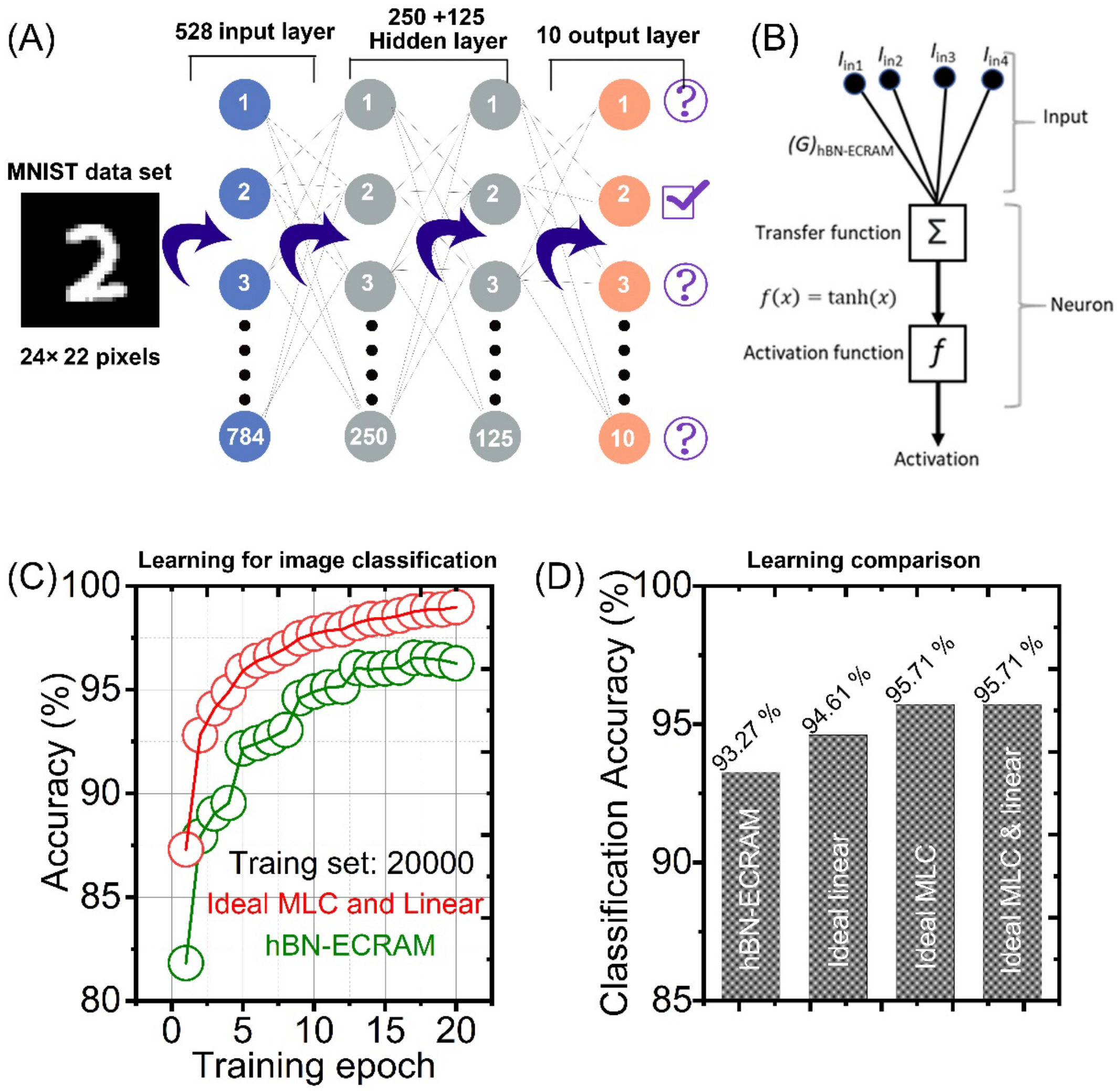

**Figure 5. MNIST data classification using hBN-ECRAM device parameters in conjunction with MLC neural network. (A)** Schematic of a three-layer MLC neural network used to classify a picture of the number "2" from digits 0 to 9. **(B)** Signal processing and transmission process in MLC neural network. **(C)** Image recognition accuracy of hBN-ECRAM and ideal device with the number of training epochs. **(D)** hBN-ECRAM accuracy comparison with ideal method.

**Table 1.** Comparison of ECRAM based artificial synaptic devices.

| $H^+$ based ECRAM synapse devices | | | | | | | | | |
|---|---|---|---|---|---|---|---|---|---|
| **Reference** | [9] | [10] | [7] | [28] | [8b] | [8a] | [5] | [29] | This work |
| **Device configuration** | 1T Lateral | 1T Lateral | 1S1T Vertical | 1T Lateral | 1T Lateral | 1T Vertical | 1T Lateral | 1T Vertical | 1T Vertical |
| **Ion transport layer** | Ionic liquid | Ion gel | Nafion | Ionic liquid | PEG | Nafion | Chitosan | silica | Single layer hBN |
| **Deposition process** | Liquid drop | Spin-coating | Drop-cast | Liquid drop | Spin-coating | Spin-coating | Drop-cast | Spin-coating | CVD and dry transfer |
| **Ion transport layer thickness** | ~ cm | ~ µm | ~ µm | ~ µm | ~ 90 nm | 300-400 nm | 100 µm | ~ µm | 0.33 nm |
| **Wafer-scale fabrication** | No | No | No | No | No | Yes | No | Yes | Yes |
| **CMOS combability** | No | No | No | No | No | No | No | No | Yes |
| **Switching layer** | $MoO_3$ | PEDOT:PSS | PEDOT:PSS | $VO_2$ | Carbon nanotube | $WO_3$ | IZO | $VO_2$ | $WO_3$ |
| **Active ion** | $H^+$ | $H^+$ | $H^+$ | $H^+$ | $H^+$ | $H^+$ | $H^+$ | $H^+$ | $H^+$ |
| **Operating pulse** | ± 2.5 V | ± 1 V | - 0.95 V, + 1.2 V | + 1.5 V, - 0.6 V | ± 5 V | ± 200 nA | ± 4 V | ± 1.5 V | ± 1 V |
| **Endurance [# pulses]** | 500 | $10^9$ | $10^8$ | $10^4$ | NA | $10^4$ | NA | NA | $10^5$ |
| **Retention [s]** | 50 s | NA | NA | 50 s | $10^3$ s | $10^3$ s | > 3 s | < 3 s | $10^3$ |
| **Operation speed** | 1 ms | 1 µs | ≥1 µs | 200 ms | 1 ms | 5 ms | 10 ms | 20 ms | < 10 ms |
| **Number of bits** | 6 | < 7 | 6 | < 7 | NA | > 7 | NA | NA | 6 |
| **Dynamic range** | 3.5 nA to 4.7 nA | 59 µS to 119 µS | 50 nS to 100 nS | 0.4 µS to 0.7 µS | 40 nS to 80 nS | 1.8 µS to 23 µS | 8.6 nA to 26 nA | NA | 8.36 µS to 10.53 µS |
| **Nonlinearity** $(\alpha_p/\alpha_d)$ | 0.19/0.13 | 1.6/0.7 | 0.95/0.92 | 2.20/2.92 | 3.04/2.92 | 1.63/1.56 | 0.45/0.23 | 0.86/1.23 | 0.90/0.90 |
| **Memory behavior** | Nonvolatile | Volatile | Volatile | Nonvolatile | Nonvolatile | Nonvolatile | Volatile | Volatile | Nonvolatile |
| **Operating temperature [°C]** | 27 | 90 | 27 | 27 | 27 | 27 | 27 | 27 | 27 |

**Note:** 1T-One transistor; 1S1T- One transistor-one selector